\documentclass{amsart}

\usepackage{amsmath}
\usepackage{amssymb}
\usepackage{dsfont} 	
\usepackage{bm}
\usepackage{mathtools}
\usepackage[hypertexnames=false]{hyperref}
\usepackage[scientific-notation=true]{siunitx}
\usepackage{enumerate}
\usepackage{color}
\usepackage[english]{babel}

\makeatletter
\newcommand{\addresseshere}{%
  \enddoc@text\let\enddoc@text\relax
}

\newtheorem{ResAim}{Research Aim}

\begin{document}

\title[Bayesian Model Selection \& Factor Analysis]
{Bayesian Constrained-Model Selection \\for Factor Analytic Modeling}

\author[C.F.W.\ Peeters]{Carel F.W.\ Peeters}
\address[Carel F.W.\ Peeters]{
Dept.\ of Epidemiology \& Biostatistics \\
VU University medical center Amsterdam \\
Amsterdam\\
The Netherlands}
\email{cf.peeters@vumc.nl}

\maketitle

My dissertation (Peeters, 2012a) revolves around Bayesian approaches towards constrained statistical inference in the factor analysis (FA) model. Two interconnected types of restricted-model selection are considered. These types have a natural connection to selection problems in the exploratory FA (EFA) and confirmatory FA (CFA) model and are termed Type I and Type II model selection. Type I constrained-model selection is taken to mean the determination of the appropriate dimensionality of a model. This type of constrained-model selection connects with EFA in the sense of selecting the optimal dimensionality of the latent vector. Type II model selection is taken to mean the determination of appropriate inequality, order or shape restrictions on the parameter space. The dissertation connects Type II constrained-model selection to CFA by focusing on the determination of linear inequality constraints as expressions of the direction and (relative) strength of factor loadings. The figures accompanying this article are taken from the slides of my Division 5 Awards Symposium Invited address at the APA 2015 Annual Convention in Toronto. These slides can be retrieved from \url{https://github.com/CFWP/ConventionTalk}.

\section*{Summary of Research}
Three research aims guide the dissertation. These research aims are motivated by the potential of the connection between Bayesian model selection (by the Bayes factor) and constrained statistical inference for factor analytic modeling. Classical approaches towards analyzing constrained hypotheses are restricted to a limited class of constraints and models. The Bayesian approach is more flexible. To utilize this flexibility, certain problems need to be overcome.

The main problem of the Bayes factor (Kass \& Raftery, 1995) for Type I model selection is that both the use of improper noninformative and proper but vague priors will yield indeterminate answers when the models to be compared are of differing dimension (Jeffreys, 1961). This is undesirable as especially under default prior choices the interpretation of the Bayes factor as a weighted likelihood ratio is warranted (Berger \& Pericchi, 2004). This problem reflects on Bayesian approaches towards EFA and the selection of the dimensionality of the latent vector. Most approaches use conjugate priors, usually in conjunction with a triangularity condition for rotational determination. Even when the conjugate priors are relatively uninformative (vague), the triangularity conditions induce strong prior information, calling into question the exploratory nature of Bayesian EFA. Research Aim 1 then intends to allow for the use of improper priors in Bayes factor computation for Type I constrained-model selection, allowing for formal dimensionality selection in EFA with all the benefits of the Bayesian machinery but without inducing strong prior information.

\begin{ResAim}
To construct a conceptually and computationally simple Bayes factor for Type I constrained-model selection that is determinate under usage of improper priors. Subsequently, this Bayes factor is to be embedded within a strategy towards a truly Bayesian EFA concerned with the selection of an optimal dimensionality for the latent vector.
\end{ResAim}

CFA seeks to incorporate theory into the factor model by imposing certain restrictions. These restrictions are meant to emulate sparse structure and are usually of the fixed-value equality kind with a focus on exclusion restrictions. Such rigidity in parameter specification may pose several problems. First, it implies a loss of information in the sense that more exclusion restrictions are usually applied than is necessary for identification of the FA model. Also, exclusion restrictions may amount to errors of omission, may make the unrealistic assumption that items are factorially pure, and may induce bias in estimates of the free parameters (Ferrando \& Lorenzo-Seva, 2000; van Prooijen \& van der Kloot, 2001). These issues are intricately connected to the well-known and widespread situation of exploratively obtained factor structures not being confirmed by CFA. Moreover, researchers in substantive fields usually have informed ideas regarding direction and magnitude of parameter effects that cannot be expressed using exclusion restrictions. What is wished for then, is expressions of factor structure, not through usage of exclusion restrictions in the matrix of factor loadings, but by the imposition of inequality constraints. This desire is formulated in the second guiding research aim.

\begin{ResAim}
To construct a conceptually and computationally simple Bayes factor for Type II constrained-model selection that is geared towards inequalities on regression-type parameters. Subsequently, this Bayes factor is to be embedded within a strategy that specifies factor analytic structure using inequality constraints rather than through exclusion restrictions.
\end{ResAim}

A gap in FA practice is that there is no unequivocal strategy for integrating EFA and CFA. While the \emph{modi operandi} are often viewed as distinct, it might be fruitful to view EFA and CFA as complementary techniques. For example, in the CFA model all attention regarding misspecification is geared towards the pre-specified pattern of factor loadings. The evaluation of model fit in CFA is then essentially the evaluation of a diffuse hypothesis (Hoyle \& Duvall, 2004), as it is unclear in case of misspecification if the pattern of loadings or the factor dimensionality is to blame. The third research aim proposes a factor analytic strategy seeking to integrate EFA and CFA in order to avoid embarking on the evaluation of diffuse hypotheses.

\begin{ResAim}
To let the provisions from Aims 1 and 2 conjoin in order to develop an integrative factor analytic strategy that articulates the complimentary nature of EFA and CFA.
\end{ResAim}

Part I of the dissertation covers Research Aims 1 and 2. The focus is on statistical and computational issues regarding formulation, prior selection, parameter estimation and model selection for the constrained Bayesian factor model. Part II comprises Research Aim 3 and purports that researchers often have competing theories that can be translated into inequality-constrained factor analytic models.

\section*{Part I: Statistical \& Computational Modeling}
\begin{figure}[t!]
\centering
  \includegraphics[scale = .7,]{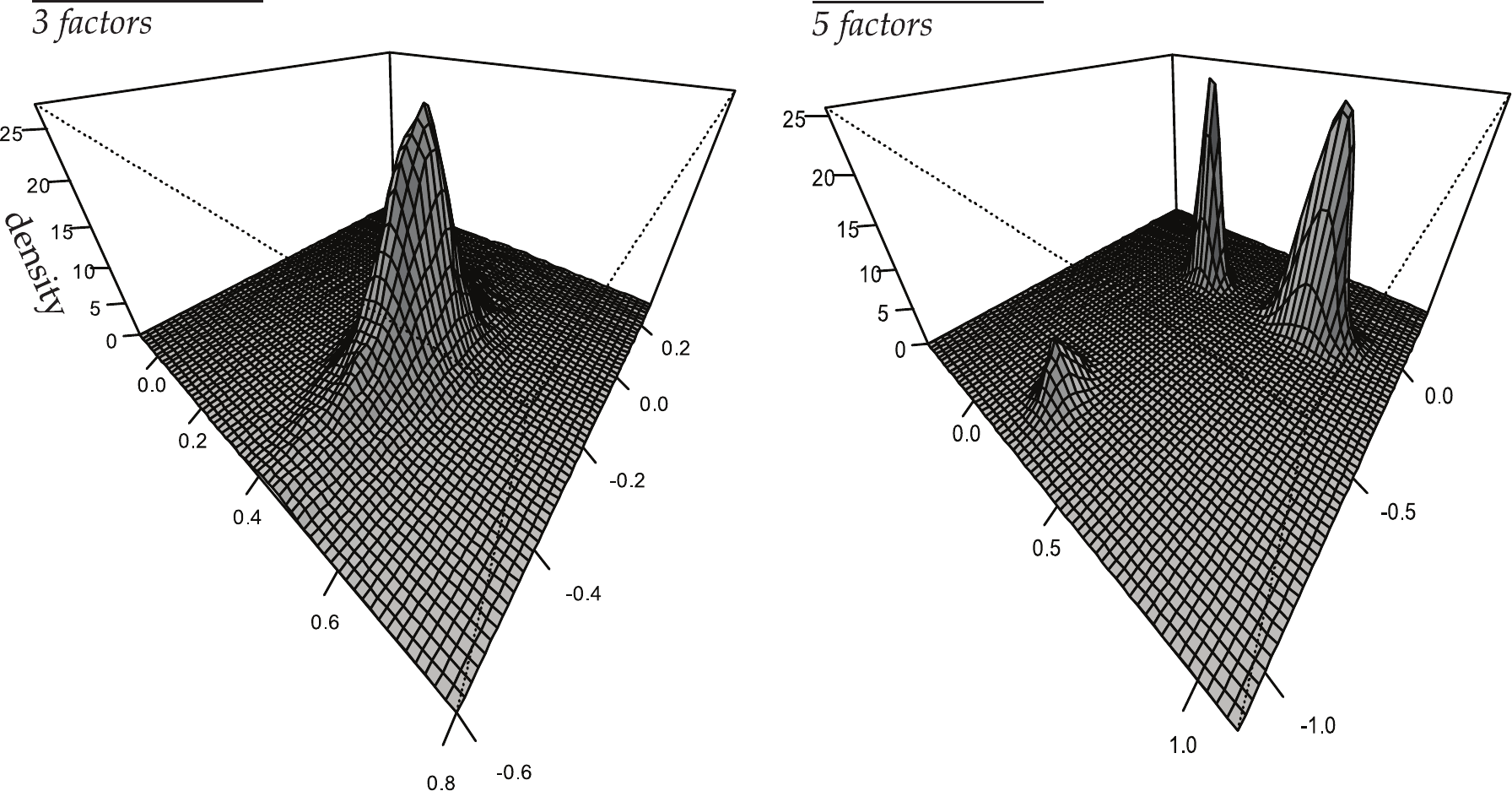}\\
  \caption{A geometric explanation of the implications of rank deficiency in the loadings matrix. The figure shows, represented by the bivariate densities on two factor loadings, what happens when one fits a 5-factor model to data generated under a 3-factor model: The (posterior) density falls apart in multiple separated regions of non-negligible posterior probability for which transition probabilities are low (or nil). This problem of rank deficiency accounts for the oft-observed fact that the likelihood-ratio test and information criteria have a tendency to overestimate model size in EFA, i.e., they tend to retain too many factors. See \url{https://github.com/CFWP/ConventionTalk} or Peeters (2012a) for additional explanation.
}\label{GEOrip}
\end{figure}

Part I starts by reviewing Markov chain Monte Carlo (MCMC) computation of the marginal likelihood. The Bayes factor consists of a ratio of marginal likelihoods. The candidate estimator method for marginal likelihood computation (Besag, 1989; Chib, 1995) is adapted to deal with (i) improper noninformative priors and (ii) the existence of (well-separated) symmetric posterior modes due to permutative invariance over the parameter indices, such that the ensuing Bayes factor is still determinate. Pending certain conditions, the provisions provide for what can be seen as a simulation consistent MCMC implementation of well-known default Bayes factors (Berger \& Pericchi, 1996). This automated candidate estimator is subsequently applied to latent factor dimensionality selection in EFA. It is shown that a failure to abide certain regularity conditions on the FA model may result in violation of a crucial regularity condition for simulation consistency of estimates stemming from MCMC sampling. This implies that Bayesian approaches towards factor analytic dimensionality selection may be afflicted by some of the same regularity conditions that hamper classical approaches (see Figure \ref{GEOrip}). The dissertation proposes an assessment strategy that ensures abidance of the regularity condition for simulation consistency such that the automated candidate estimator provides an appropriate stopping rule for factor analytic data compression. In passing, a truly Bayesian EFA is proposed.

Part I subsequently deals with a set of conditions for rotational identification of the oblique factor solution under usage of fixed zero elements in the factor loadings matrix (see also Peeters, 2012). It is shown that the well-known conditions for the oblique factor correlation structure (J\"{o}reskog, 1979) need to be amended in order to obtain global rotational uniqueness. The amended condition set provides a way to design an unrestricted solution to the (Bayesian) CFA model. Unrestricted solutions correspond to EFA: An unrestricted confirmatory factor model (UCFM) is a FA model that places only minimal restrictions on the model parameters for achieving global rotational uniqueness of the factor solution. The restrictions are (in contrast to EFA) chosen so that they convey preconceived theoretical meaning and thus eliminate the need for post-hoc rotation of the solution for interpretation purposes.

\begin{figure}[t!]
\centering
  \includegraphics[scale = .8]{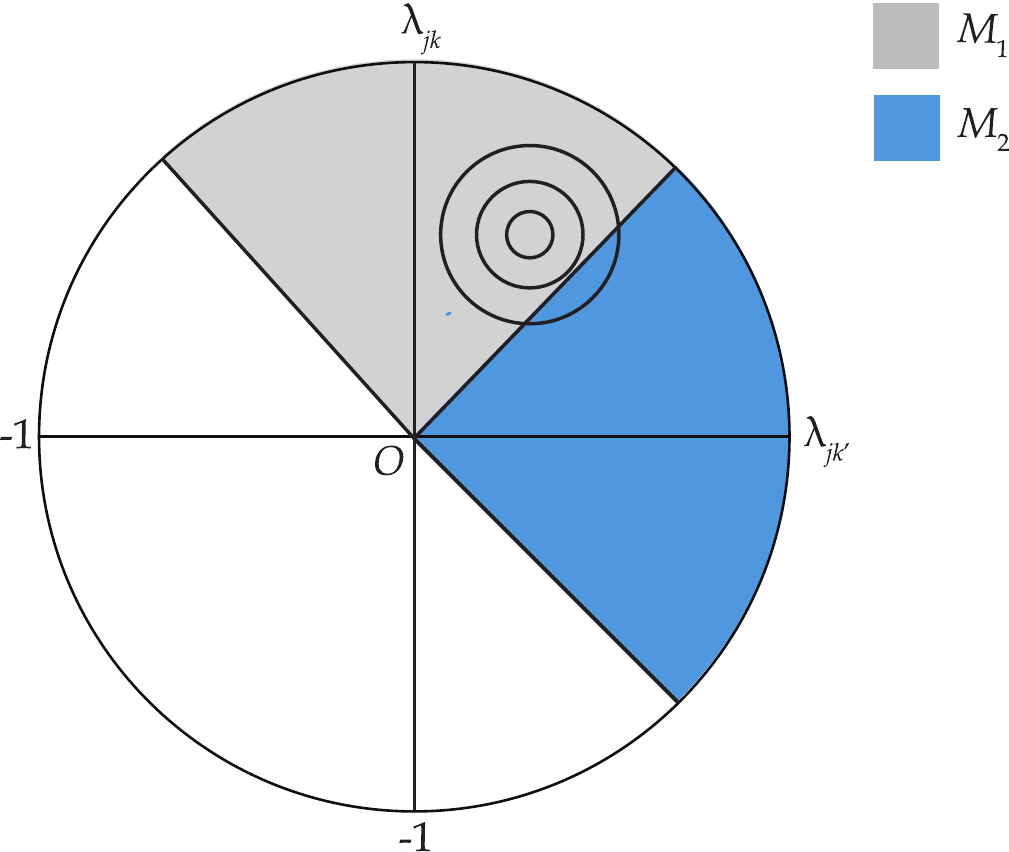}\\
  \caption{This figure contains some geometric intuition on how Bayesian inequality-constrained-model selection would work for the factor analytic model. The schematic considers a unit circle, representing the two-dimensional prior space on 2 factor loadings bound by the communality of the standardized FA solution. The concentric circles represent the (location of the) posterior probability mass. Two competing inequality-constrained models are then considered. The feasible space of Model 1 is represented  in grey. The feasible space of Model 2 is represented in blue. The Bayes factor for an inequality-constrained model to the unconstrained model boils down to the ratio of the posterior probability mass over the prior probability mass satisfying the constraints that define the model. It is then clear that Model 1 is more likely for the data at hand: Both models are of the same complexity (the feasible space comprises $1/4$ of the prior space) but the posterior mass, however, is mostly located in the feasible region defined by Model 1. See \url{https://github.com/CFWP/ConventionTalk} or Peeters (2012a) for additional explanation.
}\label{Cplexity}
\end{figure}

The UCFM is pivotal in designing a Bayesian framework that takes parameter restrictions in the context of CFA beyond exclusion restrictions, by allowing inequality and approximate equality constraints to express substantive theoretical ideas regarding direction and magnitude of effect of factor loadings. This framework first requires the development of a Bayes factor for Type II constrained-model selection. Second, a strategy is developed for the demarcation of competing inequality-constrained formulations of factor analytic correlation structure. The strategy consists of choosing as a base model a UCFM. Substantive theory is then not represented by structural exclusions to express a pre-specified loading pattern, but by imposing inequalities on and between the free parameters in the loadings matrix (see Figure \ref{Example}). It is shown that when (i) proper but noninformative priors are chosen that are flat on the parameter space of the parameters on which inequalities are placed; and (ii) all competing inequality-constrained models are subsets of the UCFM; then the ensuing Bayes factor is determinate, its complexity is well-defined, and its computation is greatly simplified. Under this framework model fit and model complexity are explicitly connected to, respectively, the posterior and prior probability mass satisfying the constraints that define the constrained model (see Figure \ref{Cplexity}).

\begin{figure}[t!]
\centering
  \begin{equation}\nonumber
  \mathbf{\Lambda}_{1}=\left[
  \begin{array}{r@{\lambda}l c r@{\lambda}l}
    &_{11}             & > & |&_{12}|   \\
    &_{21}>0           &   &  &_{22}=0 \\
    &_{31} < -.3       &   &  &_{32} > .3   \\
    &_{41}             & > & |&_{42}|   \\
    &_{51}             & > & |&_{52}|   \\
    &_{61}=0           &  & &_{62}>0 \\
  \end{array}
  \right]
  \begin{array}{l}
    \mbox{item 1}\\
    \mbox{item 2}\\
    \mbox{item 3}\\
    \mbox{item 4}\\
    \mbox{item 5}\\
    \mbox{item 6}\\
  \end{array},
\end{equation}
\\ ~
\\
vs.
\\
\begin{equation}\nonumber
  \mathbf{\Lambda}_{2}=\left[
  \begin{array}{r@{\lambda}l c r@{\lambda}l}
    |&_{11}|             & < & -&_{12}   \\
    &_{21}>0           &  & &_{22}=0 \\
    |&_{31}|             & < & &_{32}   \\
    |&_{41}|             & < & -&_{42}   \\
    &_{51}             & > & -&_{52}   \\
    &_{61}=0           &  & &_{62}>0 \\
  \end{array}
  \right]
  \begin{array}{l}
    \mbox{item 1}\\
    \mbox{item 2}\\
    \mbox{item 3}\\
    \mbox{item 4}\\
    \mbox{item 5}\\
    \mbox{item 6}\\
  \end{array}.
\end{equation}\\
  \caption{The dissertation develops methodology to select, from a batch of competing inequality-constrained factor structures, the one structure most supported by the data in terms of a balancing of model fit and model complexity. Substantive factor analytic theory is then not represented by exclusions restrictions, but by imposing (competing) systems of inequality constraints on and between the free parameters in the loadings matrix. This figure then gives a simple example of the possibilities. A situation is considered with 6 items and 2 latent factors and two competing inequality-constrained structures are given. It should be clear that the ability to directly compare these models against each other allows for higher specificity in the testing of factor structure.
}\label{Example}
\end{figure}

\section*{Part II: Applications}
Part II conjoins the developments from Part I in an alternative factor analytic strategy merging EFA and CFA. This strategy lets EFA precede inequality-constrained CFA efforts by making it part of a total inferential procedure involving the selection of an optimal factor dimension before competing confirmatory structures are assessed. The strategy consists of the following steps:

\begin{enumerate}[i.]
\item	Embark on evaluating a series of unrestricted (EFA) models with respect to their factor dimensionality;
\item	Once the latent factor dimensionality is settled, specify a UCFM;
\item	Formulate, using the UCFM as a base model, competing inequality-constrained factor structures making use of a system of inequality constraints on and between the free parameters in the loadings matrix;
\item	Compute for each constrained model the Type II constrained-model selection Bayes factor and determine the constrained model best supported by the data.
\end{enumerate}

This alternative strategy is brought to bear on research regarding the metabolic syndrome (see Chapter 5 of the dissertation and Peeters, Dziura, \& van Wesel, 2014). The conception metabolic syndrome (MBS) refers to a clustering of interrelated risk factors of metabolic origin (Unwin, 2006). MBS is thought to be a precursor for the development of type 2 diabetes mellitus and atherosclerotic cardiovascular disease. The syndrome has social, behavioral, educational and psychological components, as MBS might result from maladaptive human metabolism in the face of food energy abundance in combination with a sedentary lifestyle (Wilkin \& Voss, 2004). A high-profile MBS data set with anthropometric measurements on overweight and obese children and adolescents is reanalyzed using the alternative strategy. The alternative strategy is able to extract information with higher specificity. The findings may give inroads for behavioral and educational approaches towards MBS prevention.

\section*{Contributions and Relevance}
The contributions of the dissertation are connected to the research aims stated earlier. With the fulfillment of Aim 1 a generic computational procedure for Type I Bayes factors is available that is determinate under usage of default priors. This subsequently allows for improper prior usage in Bayesian EFA, thus respecting its exploratory nature. The developments under Aim 1 also spur learning on some lesser known indeterminacies in the factor model and their interrelationships with computational approaches towards dimensionality selection. These results also hold importance for non-Bayesian approaches towards factor analytic dimensionality selection. They imply that for informed decisions regarding factor dimensionality, likelihood ratio and information theoretic approaches benefit from a complete exploration of the likelihood, which can be achieved by objective Bayesian methods.

Fulfilling Aim 2 extends Bayesian model selection efforts regarding Type II model selection (e.g., Klugkist \& Hoijtink, 2007; Mulder, Hoijtink, \& Klugkist, 2010) and adds to the body of literature regarding inequality-constrained inference on regression-type parameters. More specifically, this research aim holds importance for the theory and technique of CFA. The formulation and development of inequality-constrained models allows for theoretically meaningful constrained coefficients beyond those needed simply to identify a model. This allows for higher specificity in model formulation as the direction and (relative) strength of factor loadings can be included in a formal model selection procedure. This may lead one to extract more information from a single analysis. The developments emphasize a break with simple structure models and may be viewed as constituting an alternative take on CFA.

Aim 3 answers the call for the integration of EFA and CFA (Steiger, 1994). The integrative strategy that is proposed connects EFA and CFA as complementary techniques that may be part of a total inferential procedure that aims to avoid evaluating diffuse model hypotheses. While Aims 1 and 2 are mostly of theoretical relevance, Aim 3 is thus of practical importance. The usefulness of the technical developments as merged in the integrative strategy is exemplified by bringing this strategy to bear on a published data set concerning risk factors for MBS. With the integrative strategy, it is shown that more and unexpected information can be extracted from these data.

\section*{Acknowledgements}
I would like to thank the APA Division 5 Award Committee for bestowing me with the 2015 Anne Anastasi Distinguished Dissertation Award. I express my gratitude towards my supervisors, Peter G.M. van der Heijden and Herbert Hoijtink. I am also indebted to the people who provided data for the applications in the dissertation: Bill Hopkins, Karin Lasthuizen, Peter Esaiasson, and James Dziura. Lastly, I acknowledge Floryt van Wesel who was, among many things, a collaborator on the application-oriented chapters of the dissertation.

This research was supported by grant NWO-VICI-453-05-002 of the Netherlands Organization for Scientific Research (NWO).
It is based on the first chapter of my unpublished PhD dissertation (2012a).
This version is a preprint of:
Peeters, C.F.W. (2016). Bayesian Constrained-Model Selection for Factor Analytic Modeling. \emph{The Score}, April Issue
(\url{http://www.apadivisions.org/division-5/publications/score/2016/04/index.aspx}).
David Herzberg's editorial help in improving this manuscript is gratefully acknowledged.

\bibliographystyle{amsplain}

\addresseshere

\end{document}